\documentclass[aps,prd,twocolumn,superscriptaddress,showpacs,floatfix]{revtex4-1} 
\usepackage{graphicx}% Include figure files
\usepackage{dcolumn}   % needed for some tables
\usepackage{bm}        % for math
\usepackage{amssymb}   % for math
\usepackage{color}
\usepackage{ifluatex}
\usepackage[utf8]{\ifluatex lua\fi inputenc}
\usepackage[T1]{fontenc}
\usepackage{newtxtext,newtxmath}
\usepackage{epstopdf}
\usepackage{slashed}
\usepackage{lineno}

\usepackage[FIGTOPCAP]{subfigure}
\usepackage{stackengine}

\begin{document}
\title{$\Xi(1690)^-$ production in the $K^-p\to K^+K^-\Lambda$ reaction process near threshold}
\author{Jung Keun Ahn}
\affiliation{Department of Physics, Korea University, Seoul 02841, Republic of Korea}
%\affiliation{Center for Extreme Nuclear Matters (CENuM), Korea University, Seoul 02841, Republic of Korea}
\author{Seung-il Nam\footnote{E-mail: {\tt sinam@pknu.ac.kr}}}
\affiliation{Department of Physics and Institute for Radiation Science \& Technology (IRST), 
Pukyong National University (PKNU), Busan 608-737, Republic of Korea}
%\affiliation{Center for Extreme Nuclear Matters (CENuM), Korea University, Seoul 02841, Republic of Korea}
\affiliation{Asia Pacific Center for Theoretical Physics (APCTP), Pohang 790-784, 
Republic of Korea}\date{\today}

\date{\today}

\begin{abstract}
We investigate $\Xi(1690)^-$ production from 
the $K^-p\to K^+K^-\Lambda$ reaction within the effective 
Lagrangian approach at the tree-level Born approximation.  
We consider the $s$- and $u$-channel $\Sigma/\Lambda$ ground states 
and resonances for the $\Xi$-pole contributions, 
in addition to the $s$-channel $\Lambda$, $u$-channel nucleon pole, 
and $t$-channel $K^-$-exchange for the $\phi$-pole contributions. 
The $\Xi$-pole includes $\Xi(1320)$, $\Xi(1535)$, $\Xi(1690)(J^p=1/2^-)$,
and $\Xi(1820)(J^p=3/2^-)$. We calculate the Dalitz plot density of 
$(d^2\sigma/dM_{K^+K^-}dM_{K^-\Lambda}$ at 4.2 GeV$/c$) 
and the total cross sections for the $K^-p\to K^+K^-\Lambda$ 
reaction near the threshold. The calculation results are 
in good agreement with previously acquired experimental data. 
Using the parameters from the fit, we present 
the total and differential cross sections for the two-body 
$K^-p\to K^+\Xi(1690)^-$ reaction near the threshold.
In our calculation, a strong enhancement 
at backward $K^+$ angles is predicted because of the dominant $u$-channel contribution. 
We also demonstrate that the Dalitz plot analysis for $p_{K^-}=1.915 -- 2.065$ GeV/c enables 
us to access direct information regarding $\Xi(1690)^-$ production, 
which can be tested by future $K^-$ beam experiments. The possible spin-parity 
states of $\Xi(1690)^-$ are briefly discussed as well.
\end{abstract}
\maketitle

\section{Introduction}

While most low-lying baryons fit primarily into SU(3) multiplets, 
the $\Xi$ spectrum is still far from being established. 
In the $S=-2$ sector, only the ground octet and decuplet states, 
$\Xi(1320)$($J^p=1/2^+$) and $\Xi(1530)$($J^p=3/2^+$), are 
well-established with four-star ratings \cite{pdg}.
Three-star states include $\Xi(1690)^-$, 
$\Xi(1820)$($J^p=3/2^-$), $\Xi(1950)$, and $\Xi(2030)$. 
The third state of $\Xi$ has not yet been confirmed between
$\Xi(1620)$ and $\Xi(1690)^-$, and theoretical model predictions are
still controversial \cite{ramos, garcia, oh, xiao, sekihara, miyahara, khem}.
The existence of $\Xi(1620)$ near $\Lambda\overline{K}$ is
dubious and requires further experimental confirmation with higher
statistics. 

$\Xi(1690)^-$ is near the $\Sigma\overline{K}$ threshold, and its 
existence has been firmly established by several experiments 
\cite{dionisi, biagi, biagi2, belle}. 
Recently, the BaBar Collaboration \cite{babar} reported that $J=1/2$ assignment was favored for 
$\Xi(1690)^-$ from its decay angular distribution. 
The $\Xi(1690)^0$ was reconstructed from $\Lambda K_S^0$ 
in the $\Lambda_c^+\to \Lambda K_S^0 K^+$ decay, taking
into account possible interference with 
$a_0(980)^+$ decaying to $K_S^0K^+$. Nevertheless, its spin and parity 
have not yet been unambiguously determined. 

$\Xi(1820)$ is the only state for which spin-parity 
is determined ($3/2^-$).  
The ordinary quark models predict that $3/2^-$ and $1/2^-$ 
should be almost degenerate, as in the case of $N^*$. 
Recall that $\Lambda(1405)$ lies near $\overline{K}N$ and it
has a large mass difference ($\approx 100$ MeV) from 
the doublet $\Lambda(1520)$. Therefore, either of $\Xi(1620)$
or $\Xi^*(1690)^-$ can be a spin partner of $\Xi(1820)$, and the rest
can be regarded as an $S=-2$ analogue state of $\Lambda(1405)$, namely 
$\Lambda \overline{K}$ or $\Sigma \overline{K}$ molecular states \cite{sekihara}.

Experimentally, $\Lambda_c^+\to\Lambda K_S^0 K+$ is particularly
attractive, as high-statistics data are available from Belle/Belle-II 
and LHCb Collaborations. However, the interference between $\Xi(1690)^-$
and $a_0(980)$ appears with a fixed crossing location in the phase space.
The phase in the interference between the two resonances could change the 
spin analysis result.  

In this respect, it is necessary to carry out 
a $\Xi(1690)^-$ production experiment using 
the $(K^-,K^+)$ reaction near the threshold. $\Xi(1690)^-$ is 
produced in the $(K^-,K^+)$ reaction and decays to $\Lambda K^-$. 
In the $K^-p\to K^+K^-\Lambda$ reaction, the $\phi(1020)\to K^+K^-$ amplitude 
could interfere with the $\Xi(1690)^-$ production amplitude. 
However, the $\phi(1020)$ resonance is very narrow, so it can readily be
isolated from the $\Xi(1690)^-$ resonance. Moreover, 
the relative location of the interference region can change with
the $K^-$ beam momentum. 

For the $K^-p\to K^+K^-\Lambda$ reaction, there have been 
no experimental efforts since the era of the bubble chamber. 
Moreover, bubble chamber data are also very limited near the threshold.
Schlein {\it et al.} \cite{schlein} 
reported the first measurement of the $K^-p\to K^+K^-\Lambda$ reaction 
using a 1.95 GeV$/c$ $K^-$ beam with a 72-$in$ hydrogen bubble chamber. 
They observed only 24 events for $K^-p\to K^+K^-\Lambda$ and 
studied the $\phi$ resonance only.   
Badier {\it et al.} studied the $K^-p\to K^+K^-\Lambda$ reaction 
using a 3 GeV$/c$ $K^-$ beam 
with an 81-cm hydrogen bubble chamber \cite{badier}. 
Because of very limited statistics, no resonances were
found in the $\Lambda K^-$ mass spectrum. Bellefon {\it et al.} \cite{bellefon} 
reported total cross sections for the $K^-p\to K^+K^-\Lambda$
reaction from 1.934 to 2.516 GeV$/c$. A total of 271 events 
were recorded in a 2-m hydrogen bubble chamber. 
The highest statistics data are available 
from the $K^-p$ experiment at 4.2 GeV$/c$, involving 2935 events from
a 2-m hydrogen bubble chamber \cite{gay}. 
The Dalitz plot for the $K^-p\to K^+K^-\Lambda$ reaction is available. 
It is therefore crucial to perform a high-statistics experiment involving 
$\Xi^\ast$ production with a high-intensity $K^-$ beam and its decay
distribution measurement 
to firmly determine their spin and parity; this type of experiment is possible at the J-PARC facility.  
   
In this paper, we report numerical calculation results for the
production of $\Xi(1690)^-$ from the $K^-p\to K^+K^-\Lambda$ reaction within
the effective Lagrangian approach. 
We consider low-lying $\Sigma/\Lambda$ resonances 
in $s$- and $u$-channels for the $\Xi$-pole contributions, and $s$-channel
$\Lambda$, $u$-channel nucleon pole, and $t$-channel $K^-$-exchange 
processes for the $\phi$-pole contributions. 
The $\Xi$-pole includes four $\Xi$ states:
$\Xi(1320)$, $\Xi(1535)$, $\Xi(1690)(J^p=1/2^-)$, 
and $\Xi(1820)(J^p=3/2^-)$. 
We calculate the total and differential cross sections for 
the $K^-p\to \Xi(1690)^-K^+$ reaction 
in a beam momentum range from 2.1 GeV$/c$ to 2.3 GeV$/c$.
We also demonstrate that 
the Dalitz plot analysis of the $K^-p\to K^+K^-\Lambda$ reaction enables
us to access direct information concerning the $\Xi(1690)^-$ production. 
The double-polarization asymmetry turns out to be essential for determining 
the spin and parity quantum numbers of $\Xi(1690)^-$ via experiments.

\section{Theoretical framework}

\begin{figure}[!h]
\includegraphics[width=7cm]{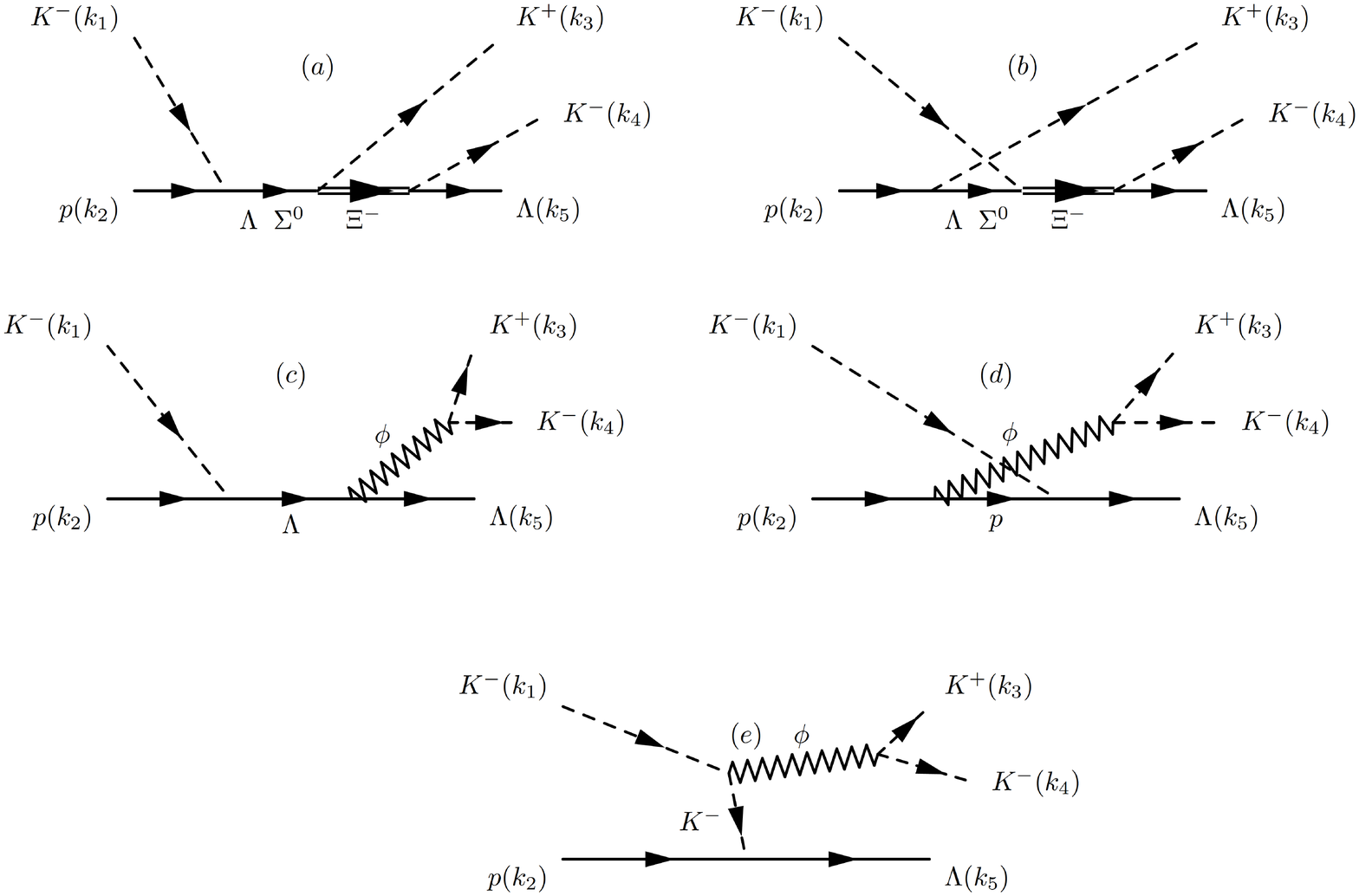}
\caption{Feynman diagrams for the $K^-p\to K^+K^-\Lambda$ reaction
at the tree-level Born approximation. 
Diagrams (a) and (b) contributed to the $\Xi$-pole, 
whereas (c), (d), and (e) contributed to the $\phi$-pole. 
The intermediate $\Lambda/\Sigma^0/\Xi^-$ denote the ground-states as well as the resonances.}       
\label{fig:diagram}
\end{figure}

In this Section, we introduce the theoretical formalism to calculate the $\Xi^*(1690)^-$
production in the $K^-p\to K^+K^-\Lambda$ reaction within the effective Lagrangian approach 
at the tree-level Born approximation. We consider five relevant Feynman diagrams for the 
$K^-p\to K^+K^-\Lambda$ reaction with the $\Xi$- and $\phi$-pole contributions, 
as shown in Fig.~\ref{fig:diagram}. 

For the $K^-p\to K^+K^-\Lambda$ reaction, the $s$- and $u$-channel diagrams 
are taken into account for the $\Xi$ production. Four $\Lambda$ states 
($\Lambda(1116)(J^p=1/2^+),~\Lambda(1405)(J^p=1/2^-),
~\Lambda(1520)(J^p=3/2^-)$, and~$\Lambda(1670)(J^p=1/2^-)$) and two
$\Sigma$ states ($\Sigma(1192)(J^p=1/2^+)$ 
and~$\Sigma(1385)(J^p=3/2^+)$) are included 
in the present calculation for $s$- and $u$-channel contributions.  
For the $\Xi$ production, four 
$\Xi$ states ($\Xi(1322)(J^p=1/2^+),~\Xi(1532)(J^p=3/2^+),
~\Xi(1690)(J^p=1/2^-)$, and $\Xi(1820)(J^p=3/2^-)$ are considered for the $\Lambda K^-$ decay channel. 
The Feynman diagrams for the $\Xi$-pole are represented in Fig.~\ref{fig:diagram}(a) and (b).  
Because the final state contains a $K^+K^-$ pair, 
the $\phi(1020)$ production can contribute to
the $K^-p\to K^+K^-\Lambda$ reaction. For the $\phi$ production, $s$-channel $\Lambda$, $u$-channel
proton, and $t$-channel $K^-$ exchange diagrams are included in the calculation,
as shown in Fig.~\ref{fig:diagram}(c), (d) and (e).   

%FIGURE============================================
%FIGURE====================================================

Here, we assume that the $\Xi(1690)^-$ has a spin-parity of $J^p=1/2^-$, 
as suggested by theoretical works~\cite{xiao, khem} 
and reported by the BaBar Collaboration \cite{babar}. 
To compute the invariant amplitudes for 
the $K^-p\to K^+K^-\Lambda$ reaction, 
we use the effective Lagrangian densities for the interaction vertices 
as follows:

%EQUATION>>>=========================================
\begin{eqnarray}
\label{eq:EFFLAG}
\mathcal{L}_{KBB}&=&-ig_{MBB}(\overline{B}{\bf\Gamma})(\gamma_5K)({\bf\Gamma}B), \\
\mathcal{L}_{K\mathcal{B}B}&=&\frac{g_{K\mathcal{B}B}}{M_K}
(\overline{\mathcal{B}}_{\mu}{\bf\Gamma}\gamma_5)
(\gamma_5\partial^\mu K)({\bf\Gamma} B), \\
\mathcal{L}_{\phi KK}&=&-ig_{KK\phi}\phi^{\mu}
\left[(\partial_{\mu}K^{\dagger})K-(\partial_{\mu}K)K^\dagger\right]+\mathrm{h.c.},
\\
\mathcal{L}_{\phi BB}&=&-g_{\phi BB}\overline{B}
\left[\gamma^\mu- \frac{\kappa_{\phi BB}}{2M_B}
\sigma_{\mu\nu}\partial^\nu \right]\phi^*_\mu B+\mathrm{h.c.}, 
\end{eqnarray}
%EQUAITON>>>====================================
where $B$ and $\mathcal{B}$ stand for 
baryons with spin-$1/2$ and spin-$3/2$, respectively. 
We should mention that, in the present calculation, we did not consider 
the $K\mathcal{B}\mathcal{B}$ vertex for brevity, as there are no experimental data available for this reaction. 

%===========================
\begin{widetext}
%TABLE>>>
\begin{center}
\begin{table}[t]
\begin{tabular}{c|cccccc}\hline
$(M,J^p)$
&$\Lambda(1116,1/2^+)$
&$\Lambda(1405,1/2^-)$
&$\Lambda(1520,3/2^-)$
&$\Lambda(1670,1/2^-)$
&$\Sigma(1193,1/2^+)$
&$\Sigma(1385,3/2^+)$\\
\hline\hline
$N(938,1/2^+)$
&$-13.24$~\cite{Rijken:1998yy}
&$0.91$~\cite{Nakayama:2006ty}
&$-10.9$0~\cite{pdg}
&$0.30$~\cite{Shyam:2011ys}
&$3.58$~\cite{Rijken:1998yy}
&$-3.22$~\cite{Nakayama:2006ty}\\
\hline
$\Xi(1322,1/2^+)$
&$3.52$~\cite{Rijken:1998yy}
&$0.91$~\cite{Nakayama:2006ty}
&$3.27$~\cite{Nakayama:2006ty}
&$-0.18$~\cite{Shyam:2011ys}
&$-13.26$~\cite{Rijken:1998yy}
&$-3.22$~\cite{Nakayama:2006ty}\\
\hline
$\Xi(1532,3/2^+)$
&$4.08$
&$-$
&$-$
&$-$
&$3.22$
&$-$\\
\hline
$\Xi(1690,1/2^-)$
&$-0.3$~\cite{khem}
&$-$
&$-$
&$-$
&$1.8$~\cite{khem}
&$-$\\
\hline
$\Xi(1820,3/2^-)$
&$6.10$~\cite{xiao}
&$-$
&$-$
&$-$
&$8.00$~\cite{xiao}
&$-$\\ \hline
\end{tabular}
\caption{Coupling constants for $KBB$ and $KB\mathcal{B}$ vertices in the present calculation are taken  
from Refs.~\cite{khem,Nakayama:2006ty,
Shyam:2011ys,Rijken:1998yy,pdg,xiao}.}
\label{TAB0}
\end{table}
\end{center}
\end{widetext}
%TABLE>>>===========================================================

We define ${\bf\Gamma}$, which depends on the parity $P$ of 
the neighboring baryon in the above interaction Lagrangian densities, 
i.e., $({\bf\Gamma} B)$ and $({\bf\Gamma}\mathcal{B})$ for instance, as follows:

%EQUATION>>>==========================================
\begin{equation}
\label{eq:}
{\bf\Gamma}=\Big\{
\begin{array}{c}
\bm{1}_{4\times4}\,\,\,\,\mathrm{for}\,\,\,\,P_{B,\mathcal{B}}=+1\\
\gamma_5\,\,\,\,\mathrm{for}\,\,\,\,P_{B,\mathcal{B}}=-1.
\end{array}
\end{equation}
%EQUAITON>>>============================================

The calculation of the invariant amplitudes is strightforward:
\begin{widetext}
%EQUATION>>>
\begin{eqnarray}
\label{eq:AMP}
%----------------------
i\mathcal{M}^{\Xi_1Y_1}_s&=&-g_{K\Lambda\Xi_1}g_{KY_1\Xi_1}g_{KNY_1}F_\Xi(s)
%\cr 
\times %&{}&\hspace{-0.4cm}
\frac{\overline{u}_\Lambda\gamma_5{\bf\Gamma}_{\Xi_1}(\slashed{q}_{4+5}+M_{\Xi_1}){\bf\Gamma}_{\Xi_1}\gamma_5{\bf\Gamma}_{Y_1}
(\slashed{q}_{1+2}+M_{Y_1}){\bf\Gamma}_{Y_1}\gamma_5u_N}
{[M^2_{K^-\Lambda}-M^2_{\Xi_1}+i\Gamma_{\Xi_1}M_{\Xi_1}][s-M^2_{Y_1}+i\Gamma_{Y_1}M_{Y_1}]},
\\
%----------------------
i\mathcal{M}^{\Xi_1Y_1}_u&=&-g_{K\Lambda\Xi_1}g_{KY_1\Xi_1}g_{KNY_1}F_\Xi(u)
%\cr 
\times %&{}&\hspace{-0.4cm}
\frac{\overline{u}_\Lambda\gamma_5{\bf\Gamma}_{\Xi_1}(\slashed{q}_{4+5}+M_{\Xi_1}){\bf\Gamma}_{\Xi_1}\gamma_5{\bf\Gamma}_{Y_1}
(\slashed{q}_{2-3}+M_{Y_1}){\bf\Gamma}_{Y_1}\gamma_5u_N}
{[M^2_{K^-\Lambda}-M^2_{\Xi_1}+i\Gamma_{\Xi_1}M_{\Xi_1}][u-M^2_{Y_1}+i\Gamma_{Y_1}M_{Y_1}]},
\\
%----------------------
i\mathcal{M}^{\Xi_3Y_1}_s&=&\frac{g_{K\Lambda\Xi_3}g_{K\Xi_3Y_1}g_{KNY_1}
F_\Xi(s)}{s}
%\cr 
\times %&{}&\hspace{-0.4cm}
\frac{\overline{u}_\Lambda{\bf\Gamma}_{\Xi_3}(\slashed{q}_{4+5}+M_{\Xi_3})
{\bf\Gamma}_{\Xi_3}{\bf\Gamma}_{Y_1}
(\slashed{q}_{1+2}+M_{Y_1})(k_3\cdot k_4){\bf\Gamma}_{Y_1}\gamma_5u_N}
{[M^2_{K^-\Lambda}-M^2_{\Xi_3}+i\Gamma_{\Xi_3}M_{\Xi_3}][s-M^2_{Y_1}+i\Gamma_{Y_1}M_{Y_1}]},
\\
%----------------------
i\mathcal{M}^{\Xi_3Y_1}_u&=&\frac{g_{K\Lambda\Xi_3}g_{K\Xi_3Y_1}g_{KNY_1}
F_\Xi(u)}{s}
%\cr 
\times %&{}&\hspace{-0.4cm}
\frac{\overline{u}_\Lambda{\bf\Gamma}_{\Xi_3}(\slashed{q}_{4+5}+M_{\Xi_3}){\bf\Gamma}_{\Xi_3}{\bf\Gamma}_{Y_1}
(\slashed{q}_{2-3}+M_{Y_1})(k_1\cdot k_4){\bf\Gamma}_{Y_1}\gamma_5u_N}
{[M^2_{K^-\Lambda}-M^2_{\Xi_3}+i\Gamma_{\Xi_3}M_{\Xi_3}][u-M^2_{Y_1}+i\Gamma_{Y_1}M_{Y_1}]},
\\
%----------------------
i\mathcal{M}^{\phi}_s&=&g_{\phi KK}g_{\phi \Lambda \Lambda}g_{KN \Lambda}F_\phi(s)
%\cr 
\times %&{}&\hspace{-0.4cm}
\frac{\overline{u}_\Lambda\left[\slashed{q}_{3-4}\right]
(\slashed{q}_{1+2}+M_Y)\gamma_5u_p}
{[M^2_{K^+K^-}-M^2_\phi+i\Gamma_\phi M_\phi][s-M^2_Y+i\Gamma_YM_Y]},
\\
%----------------------
i\mathcal{M}^{\phi}_u&=&-g_{\phi KK}g_{KNY}g_{\phi NN}F_\phi(u)
%\cr 
\times %&{}&\hspace{-0.4cm}
\frac{\overline{u}_\Lambda\gamma_5
(\slashed{q}_{5-1}+M_N)\slashed{q}_{3-4}u_p}
{[M^2_{K^+K^-}-M^2_\phi+i\Gamma_\phi M_\phi][u-M^2_N]},
\\
%----------------------
i\mathcal{M}^{\phi}_t&=&g^2_{\phi KK}g_{KN\Lambda}F_\phi(t)
%\cr 
\times %&{}&\hspace{-0.4cm}
\frac{2(k_1\cdot q_{3-4})\overline{u}_\Lambda\gamma_5u_p}
{[M^2_{K^+K^-}-M^2_\phi+i\Gamma_\phi M_\phi][t-M^2_K]},
\end{eqnarray}
%EQUAITON>>>
\end{widetext}
%=======================================================
where $M_{K^+K^-}\equiv(k_3+k_4)^2$ and $M_{K^-\Lambda}\equiv(k_4+k_5)^2$
denote the invariant masses. In the present calculation, the four-momenta of $K^-$ beam, target $p$, outgoing $K^+$, 
outgoing $K^-$, and $\Lambda$ are denoted $k_1$, $k_2$, $k_3$, $k_4$, and $k_5$, respectively, 
as shown in Fig.~\ref{fig:diagram}, while $q_{i\pm j}=k_i\pm k_j$ are the relative four-momenta for two particles, 
where $i$ and $j$ range from 1 to 5.

The coupling constants for the ground-state hadron vertices, such as $g_{KN\Lambda(1116)}$, are taken 
from the prediction of the Nijmegen soft-core potential model (NSC97a)~\cite{Rijken:1998yy}. 
The coupling constants for the $s$-wave resonances, $\Lambda(1405)$ and $\Lambda(1670)$, 
are obtained from the chiral unitary model~\cite{Nakayama:2006ty}, where the resonances are generated dynamically 
by the coupled-channel method with the Weinberg--Tomozawa (WT) chiral interaction. 
The couplings for $\Xi(1690)$ and $\Xi(1820)$ are estimated by ChUM~\cite{khem} and 
the SU(6) relativistic quark model~\cite{xiao}, respectively. 

Regarding the coupling constants with two hyperon resonances, 
such as $g_{K\Lambda^*\Xi^*}$ and $g_{K\Sigma^*\Xi^*}$, 
there is no experimental nor theoretical information available. Furthermore, it is also difficult and uncertain 
to simply employ the flavor SU(3)-symmetry relation, which is used to obtain $g_{K\Lambda^*\Xi }$ and 
$g_{K\Sigma^*\Xi}$ as in Ref.~\cite{Shyam:2011ys}. Hence, we set those coupling constants to be zero for simplicity, 
although in practice their unknown contributions can be absorbed into the cutoff parameters of the form factors. 
The strong coupling constants used in the present calculation are listed in Table~\ref{TAB0}. 

The full decay widths for $\Xi^\ast$ resonances are given as $\Gamma_{\Xi(1532)}=9.1$ MeV~\cite{pdg}, 
$\Gamma_{\Xi(1690)}=6$ MeV~\cite{khem}, and $\Gamma_{\Xi(1820)}=24$ MeV~\cite{pdg,xiao}. 
In Eq.~(\ref{eq:AMP}), we introduce the phenomenological form factors 
for the $\Xi$- and $\phi$-pole contributions to take their spatial distributions into account:
%EQUATION>>>
\begin{equation}
\label{eq:FF}
F_{\Xi,\phi}(x)=\frac{\Lambda^4_{\Xi,\phi}}
{\Lambda^4_{\Xi,\phi}+(x-M^2_x)^2},~~\mathrm{for}\,~~x=(s,u,t),
\end{equation}
%EQUAITON<<<=====================================
where $s$, $t$, and $u$ are the Lorentz-invariant Mandelstam variables. In the present calculation,
the cutoff parameters are determined to be $\Lambda_{\Xi(1322)}=1.3$ GeV, 
$\Lambda_{\Xi(1532)}=1.3$ GeV, $\Lambda_{\Xi(1690)}=0.75$ GeV, $\Lambda_{\Xi(1820)}=1.1$ GeV,  and $\Lambda_\phi=0.44$ GeV 
to reproduce the experimental data, which will be discussed in the next Section. 
We also choose the phenomenological phase factors, $e^{3i\pi/2}$ and $e^{i\pi/2}$ for the amplitudes with
the spin-$1/2$ and spin-$3/2$ $\Xi$ hyperons, respectively, as follows:

\begin{eqnarray}
i\mathcal{M}_{\rm total}=ie^{i\pi/2}\mathcal{M}_{\Xi_{3/2}}+ie^{i3\pi/2}\mathcal{M}_{\Xi_{1/2}}+i\mathcal{M}_{\phi}
\end{eqnarray} 

%%%%%%%%%%%%%%%%%%%%%%%%%%%%%%%%%%%%%%%%%%%

\section{Numerical Results for the $K^-p\to K^+K^-\Lambda$ Reaction}

\begin{figure}[!h]
\centering
\topinset{(a)}{
\hspace{0.5cm}\includegraphics[width=8.2cm]{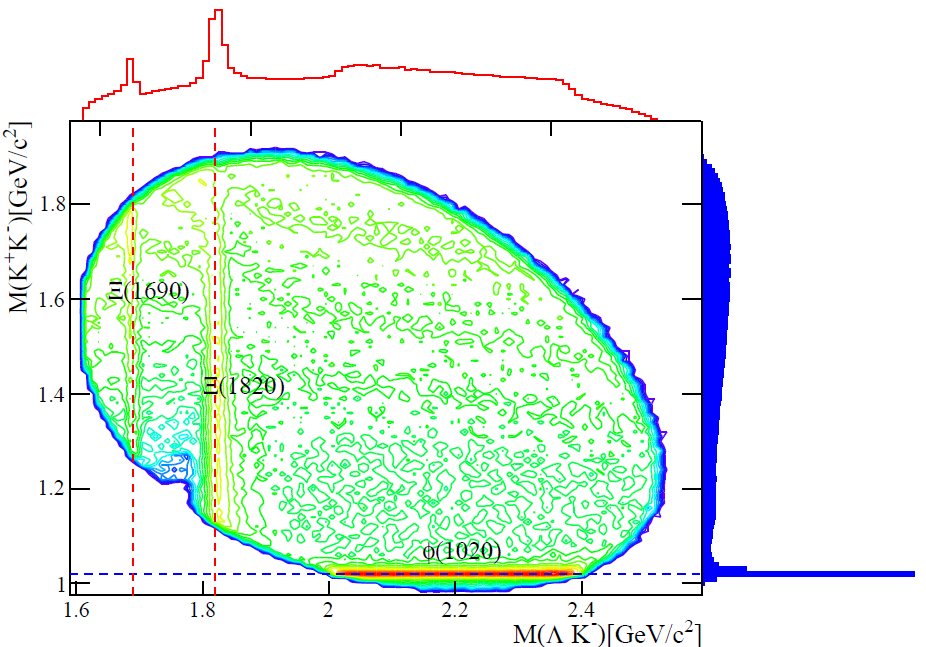}
}{0.5cm}{1.8cm}\\
\topinset{(b)}{\hspace{-1.3cm}
\includegraphics[width=6.7cm]{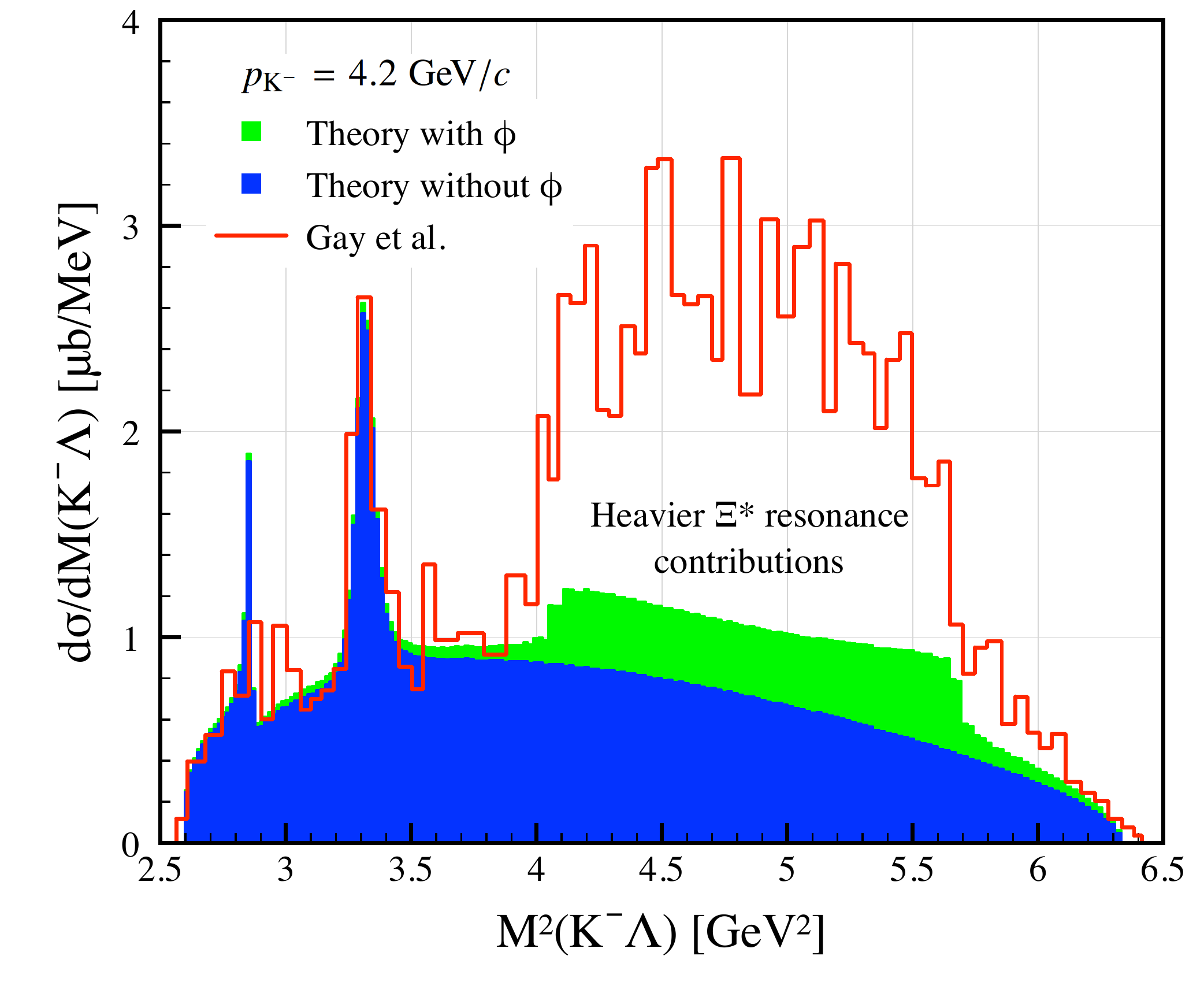}
}{0.5cm}{1.8cm}
\caption{(Color online) (a) Calculated Dalitz plot density ($d^2\sigma/dM_{K^+K^-}dM_{K^-\Lambda}$) 
for the $K^-p\to K^+K^-\Lambda$ reaction at $p_{K^-}=4.2$ GeV. 
(b) Differential cross section $d\sigma/dM_{K^-\Lambda}$ 
as a function of the invariant mass squared $M^2_{K^-\Lambda}$ 
at $p_{K^-}=4.2$ GeV. 
The green and blue areas indicate the results 
with and without the $\phi(1020)$ contribution, respectively. 
The experimental data \cite{gay} are overlaid as a histogram.}
\label{fig:gay}
\end{figure}

In this Section, we discuss the numerical results for the $\Xi(1690)$ production. We first show 
the numerical results for the $K^-p\to K^+K^-\Lambda$ reaction. 
The calculated Dalitz plot for the double differential cross section 
$d^2\sigma/dM_{K^+K^-}dM_{K^-\Lambda}$ at $p_{K^-}=4.2$ GeV$/c$ ($E_\mathrm{cm}=3.01$ GeV) 
is represented in Fig.~\ref{fig:gay}(a), 
where the $\Xi^*(1690)$ and $\Xi(1820)$ resonances appear as vertical bands, while
$\phi(1020)$ appears as a horizontal band in the bottom side. At this energy, there is no interference effect 
between $\Xi^*$s and $\phi(1020)$. 

%FIGURE====================================================
%FIGURE==
 
The Dalitz plot was projected on the $K^-\Lambda$ mass axis, as shown in Fig.~\ref{fig:gay}(b). 
The experimental data are taken from Ref.~\cite{gay}, which is the only data set available so far 
for the $K^-p\to K^+K^-\Lambda$ reaction. The experiment was performed using the $K^-$ beam at 4.2 GeV$/c$ to study 
$\Xi(1820)$ and higher resonances. 
We then fit the data with the line shape of 
our calculation result in the low-mass region below $M^2_{K^-\Lambda}=3.3\,\mathrm{GeV}^2/c^4$. 
The first bump structure near the threshold is due to the $\Xi(1690)$ production, 
providing us with information on the cutoff parameters for the form factors, 
as given in the previous Section. The green and blue areas indicate the calculation results 
with and without the $\phi(1020)$ contribution, respectively. The mass range between 4.0 and 5.7 GeV$^2/c^4$ for
the large bump structure is consistent with the $\phi(1020)$ band crossing the limited
phase space in the Dalitz plot. It should be noted that high-mass resonances decaying to $K^-K^+$ cannot 
account for the bump structure in that mass range only. In the present calculation, high-mass $K^-K^+$     
resonance like $f^\prime_2(1525)(J^p=2^+)$ is not included. Higher-mass $\Xi^*$ resonances could contribute to
the bump structure.

%FIGURE======================================================
\begin{figure}[!h]
\centering
\topinset{(a)}{
\includegraphics[width=6.5cm]{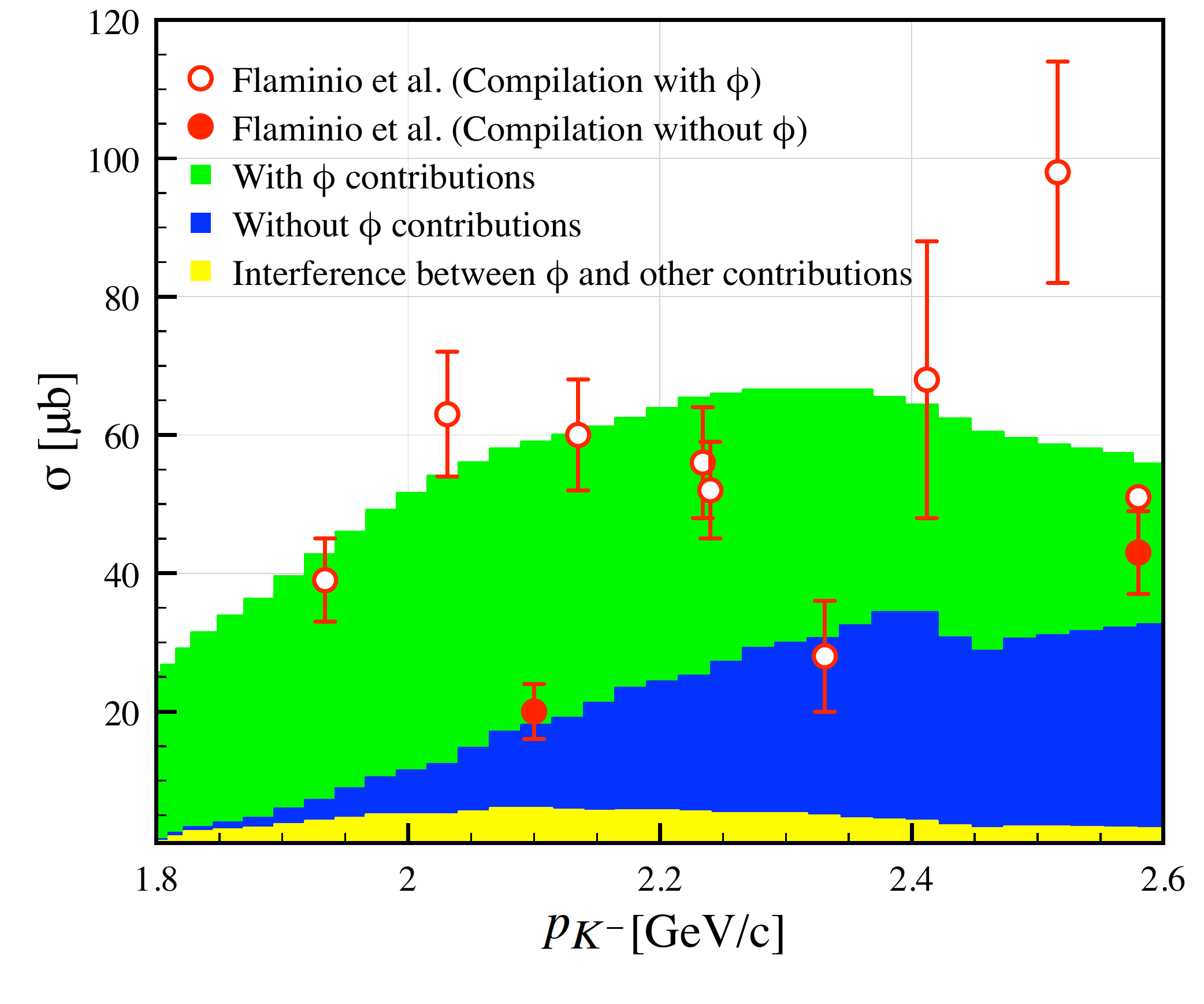}
}{0.5cm}{1.8cm}\hspace{-0.4cm}
\topinset{(b)}{
\includegraphics[width=6.5cm]{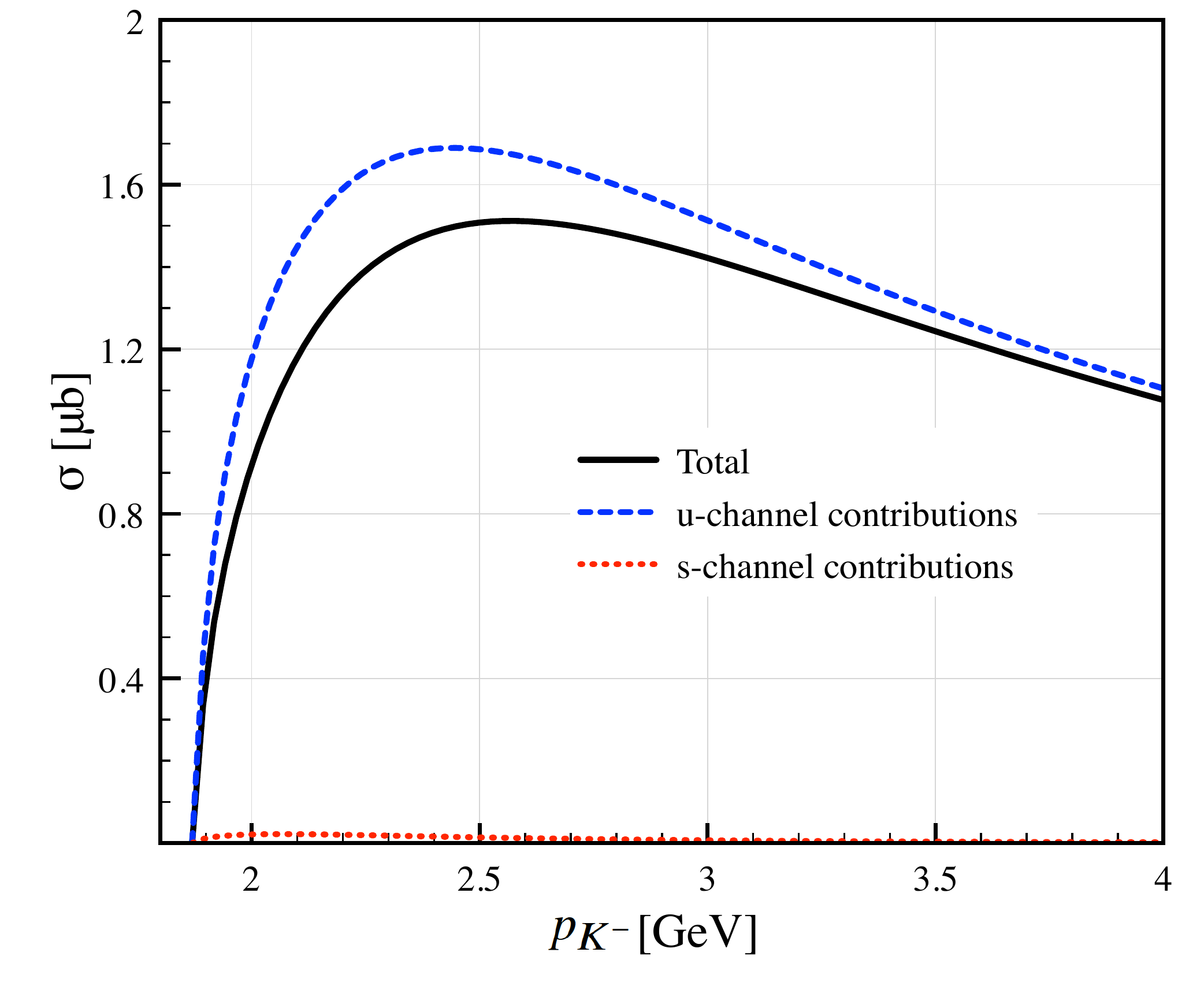} 
}{0.5cm}{1.8cm}\hspace{-0.4cm}
\topinset{(c)}{
\includegraphics[width=6.5cm]{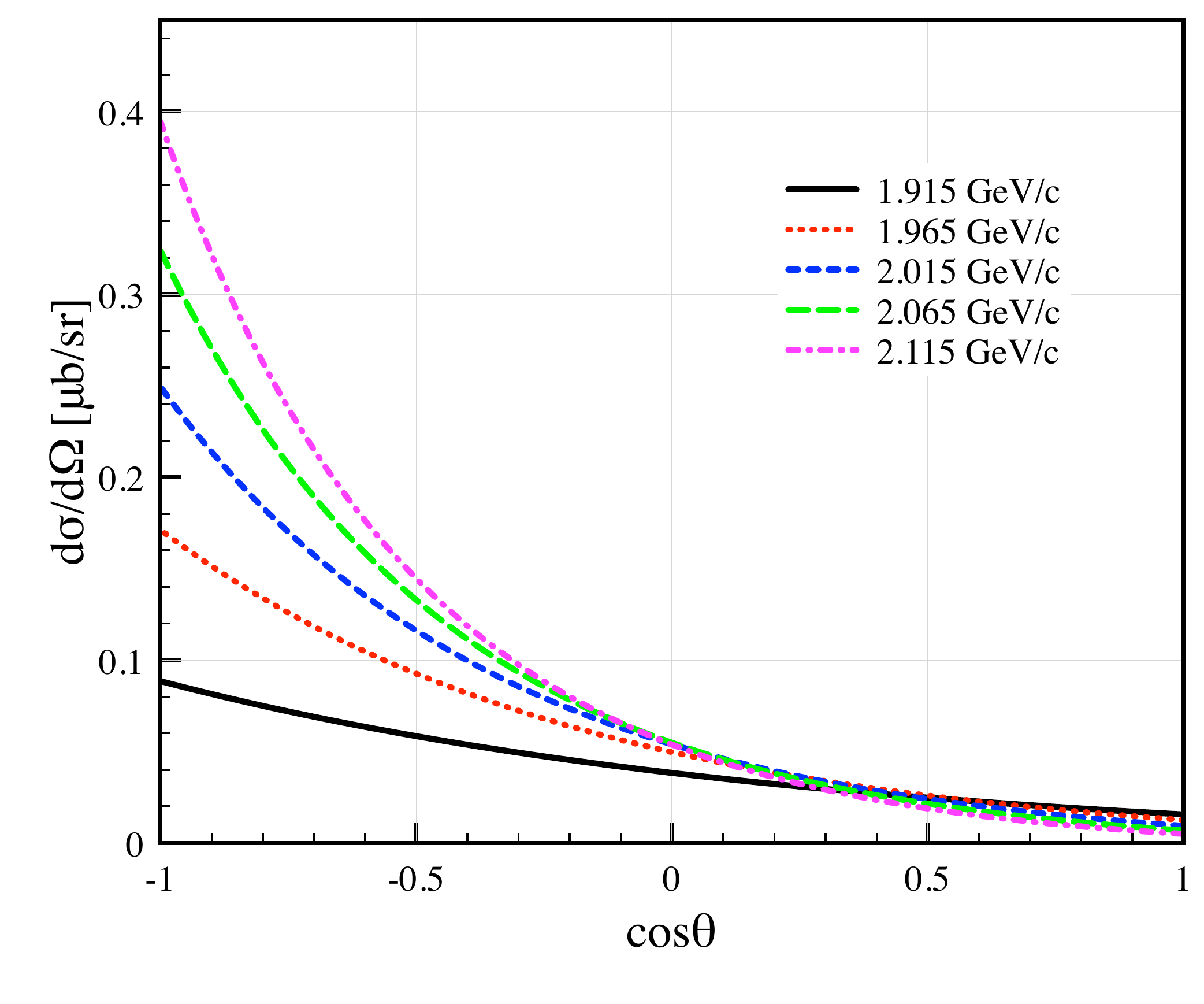}
}{0.5cm}{1.8cm}
\caption{(Color online) 
(a) Total cross section for $K^-p\to K^+K^-\Lambda$ with (green) 
and without (blue) $\phi(1020)$ contribution as functions of $p_{K^-}$. The yellow area 
at the bottom indicates interference between the $\phi$ and other contributions. 
The experimental data are taken from Ref.~\cite{Flaminio:1983fw}.  
(b) Total cross section for $K^-p\to K^+\Xi(1690)^-$ 
as functions of $p_{K^-}$. (c) Differential cross section 
for $K^-p\to K^+\Xi(1690)^-$ as functions of the angle for the outgoing $K^+$ in the c.m. frame for several beam momenta $p_{K^-}$.}       
\label{fig:xross}
\end{figure}
%FIGURE=====================================================

%FIGURE=====================================================
\begin{figure}[t]
\includegraphics[width=8cm]{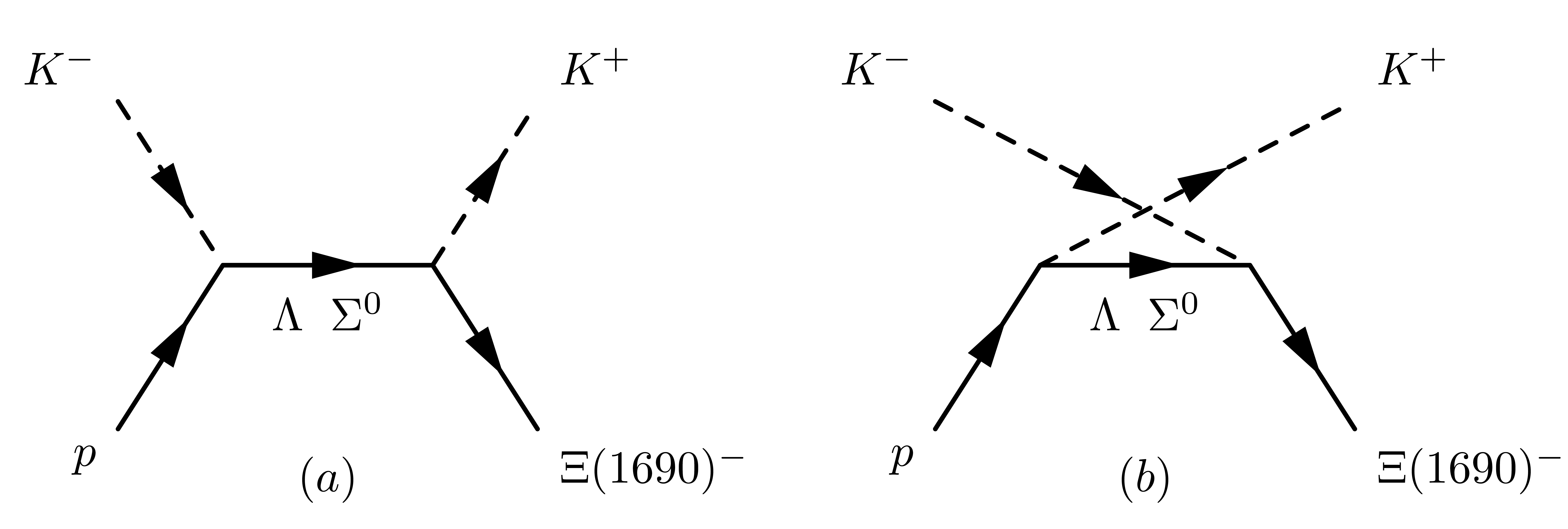}
\caption{Relevant Feynman diagrams for $K^-p\to K^+\Xi(1690)^-$.}       
\label{fig:TBP}
\end{figure}
%FIGURE=====================================================
Using the same cutoff parameters, we compute the total cross sections for the $K^-p\to K^+K^-\Lambda$
reaction. The calculation results with (green area) and without (blue area) the $\phi(1020)$ contribution 
are compared with the world data taken from Ref.~\cite{Flaminio:1983fw}, as shown in Fig. \ref{fig:xross}(a). 
The yellow area at the bottom indicates interference between the $\phi$ and other contributions. 
It turns out that our theoretical model describes the experimental data for the $K^-p\to K^+K^-\Lambda$ reaction 
qualitatively well. Enhanced $K^-\Lambda$ production between $M^2_{K^-\Lambda}=4.0$ GeV$^2/c^4$ and 
$5.7$ GeV$^2/c^4$ could be associated with a contribution from higher-mass hyperon resonances. However, we 
did not include those high-mass resonances in our present calculation, 
as the mass region is far 
beyond the $\Lambda K^-$ threshold.   

For the two-body $K^-p\to K^+\Xi(1690)^-$ reaction, we computed the $s$-channel and $u$-channel diagrams in 
Fig.~\ref{fig:TBP} with the same theoretical framework and the same parameters used for the $K^-p\to K^+K^-\Lambda$
reaction. The total cross sections are represented as a function of $K^-$ beam momentum ($p_{K^-}$) from threshold
to 4 GeV$/c$ in Fig.~\ref{fig:xross}(b). 
It increases rapidly from the threshold and peaks 
at $p_{K^-}=2.6$ GeV$/c$ ($E_{\rm cm}=2.47$ GeV) with $1.5~\mu$b, 
after which it decreases 
smoothly. As shown in Fig.~\ref{fig:xross}(b), the $u$-channel contribution is much larger than 
the $s$-channel contribution. In our present calculation, 
we set the coupling constant $(g_{KY^*\Xi^*})$ to zero to avoid further theoretical uncertainty.   
Shyam {\it et al.} ~\cite{Shyam:2011ys} assumed that 
$g_{KY^*\Xi}=g_{KY^*N}$ for the $K^-p\to K^+\Xi^-$ reaction. 
However, there is no firmly established theoretical basis for 
the coupling constants $(g_{KY^*\Xi^*})$. 
The $u$-channel hyperon propagator and form factors also provide much larger strengths 
than that for the $s$-channel one, as previously shown in Ref.~\cite{Nam:2004ub}. 

%%%%%%%%%%%%%%%%%%%%%%%%%%%%%%
\begin{figure}[!h]
\centering
\topinset{(a)}{
\includegraphics[width=6.5cm]{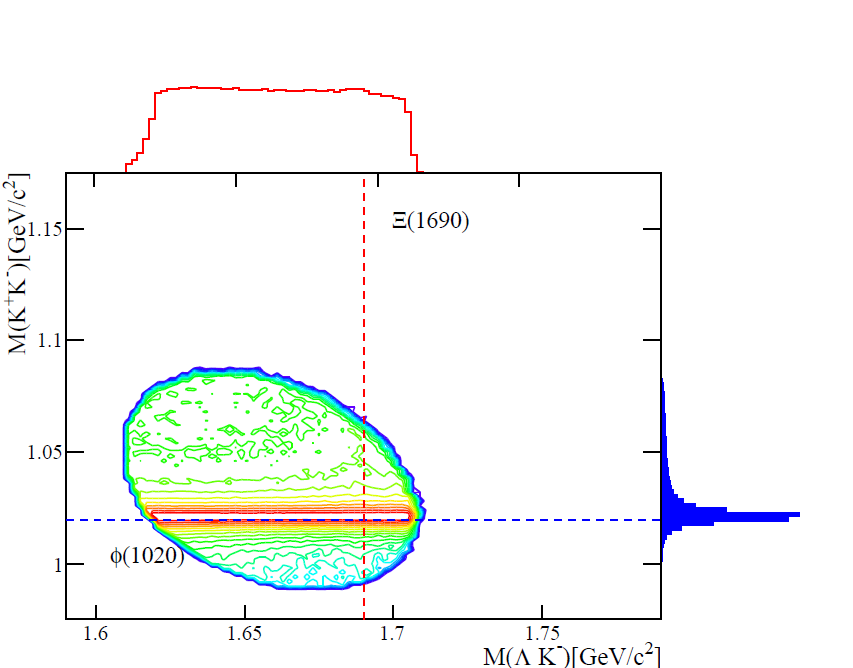}
}{0.5cm}{1.3cm}\hspace{-0.4cm}
\topinset{(b)}{
\includegraphics[width=6.5cm]{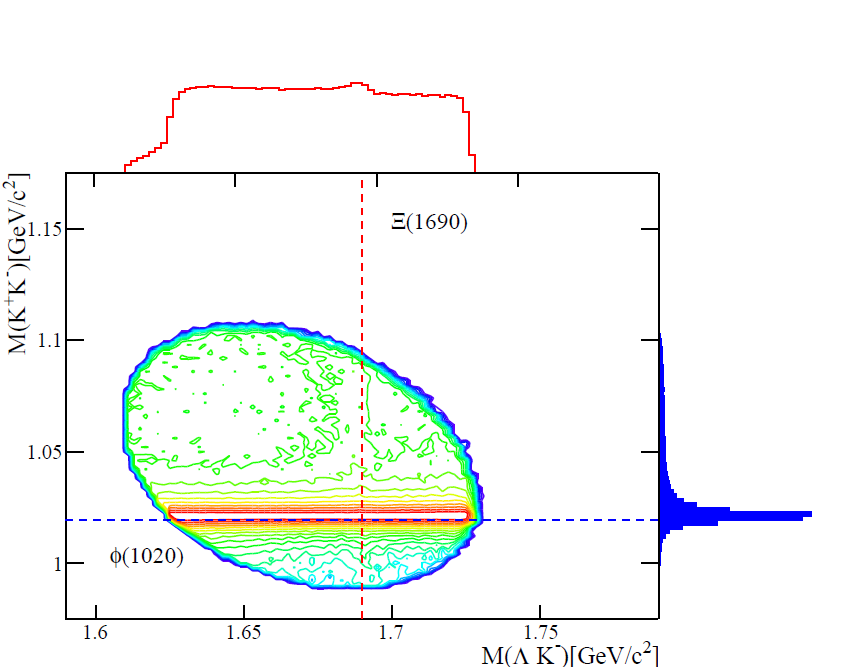}
}{0.5cm}{1.3cm}\hspace{-0.4cm}
\topinset{(c)}{
\includegraphics[width=6.5cm]{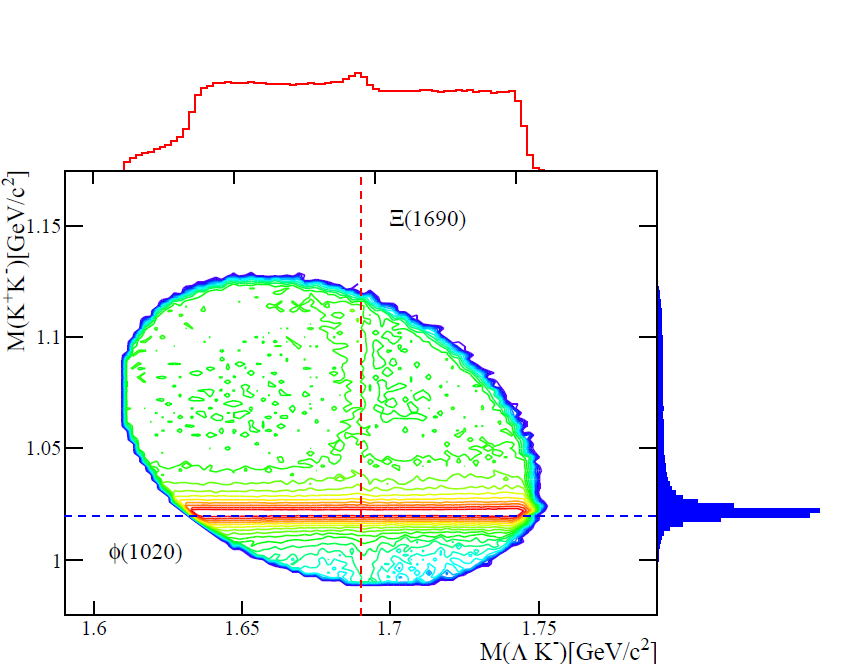}
}{0.5cm}{1.3cm}\hspace{-0.4cm}
\topinset{(d)}{
\includegraphics[width=6.5cm]{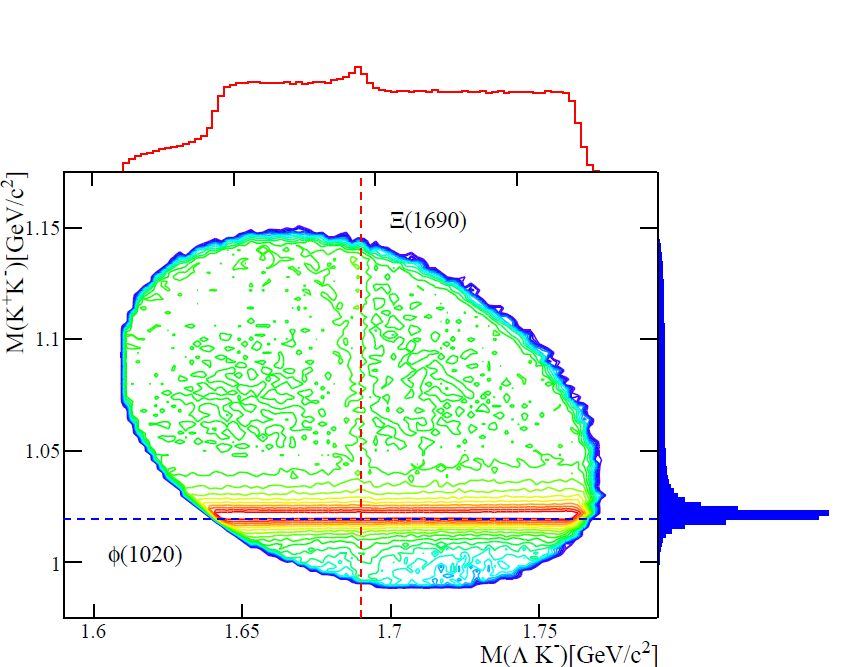}
}{0.5cm}{1.3cm}
\caption{(Color online) Dalitz plots for the $K^-p\to K^+K^-\Lambda$ reaction 
for (a) $p^{\rm lab}_{K^-}=1.915$ GeV$/c$, (b) $1.965$ GeV$/c$, (c) $2.015$ GeV$/c$ and (d) $2.065$ GeV$/c$, 
respectively. The Dalitz plots are projected on to the $\Lambda K^-$ and the $K^+K^-$ mass axes
and plotted as histograms on the top and right sides, respectively.}       
\label{fig:dalitz}
\end{figure}
%%%%%%%%%%%%%%%%%%%%%%%%%%%

The differential cross sections $d\sigma/d\Omega$ for the $K^-p\to K^+\Xi(1690)^-$ reaction 
are calculated as a function of $\cos\theta$ in Fig. \ref{fig:xross}(c), 
where $\theta$ stands for the scattering angle of the 
outgoing $K^+$ in the center-of-mass (c.m.) frame. Because of the strong $u$-channel contributions, 
one observes backward-enhanced angular distributions for the various $p_{K^-}$ values, as the energy increases. 
This backward-peaking behavior is a general feature for the double-charge 
and double-strangeness exchange $(K^-,K^+)$ process.

%============================================================================

The threshold beam momentum for the $K^-p\to K^+K^-\Lambda$ reaction is 1.687 GeV$/c$, while that for
the $K^-p\to K^+\Xi(1690)^-$ reaction 
is 1.878 GeV$/c$, which is accessible using the J-PARC Hadron-Hall Collaboration. 
The experiment for the $K^-p\to K^+K^-\Lambda$ reaction 
near the threshold
can be performed with the Hyperon Spectrometer \cite{ahn} at the K1.8 beam line of J-PARC. 
One can measure all the charged particles not only from the $\Xi(1690)^-\to K^-\Lambda$ decay, but also
$\Sigma^-K^0$, $\Sigma^0 K^-$, $\Xi^-\pi^0$, and $\Xi^0\pi^-$ decays. All the $\Xi(1690)^-$ decay modes  
contain three charged particles with one missing neutral particle in some channels, which enables us to
reconstruct $\Xi(1690)^-$ without any kinematical ambiguity. 

The sizable cross sections of a few $\mu$b for the $K^-p\to K^+\Xi(1690)^-$ reaction also encourage future experiments
using a high-intensity $K^-$ beam \cite{ryu}. 
According to the calculated Dalitz plot density, simulated events for the $K^-p\to K^+K^-\Lambda$ reaction are generated
over the phase space available. We assume a uniform experimental acceptance for the $K^+K^-\Lambda$ phase space. 
The Dalitz plots for the $K^-p\to K^+K^-\Lambda$ reaction are plotted 
in Fig. \ref{fig:dalitz} for four different $K^-$ beam momenta, from 1.915 GeV$/c$ to 2.065 GeV$/c$. 
Because the $\phi(1020)$ production is predominant, 
it is difficult to identify $\Xi(1690)^-$ in the $\Lambda K^-$ mass distribution without the $\phi$-band exclusion. 
The $\phi$ band is so narrow that we can remove the $\phi$ events by excluding the $K^+K^-$ mass band for the $\phi$.   
The crossing points between the $\Xi(1690)^-$ and the $\phi(1020)$ resonances change with the $K^-$ beam momentum. 
This enables us to study $\Xi(1690)^-$ in various kinematical regions, where the interference effects with 
the $\phi(1020)$ resonance are different.  
  
%FIGURE-===========================================================================
 
\begin{figure}[!hpbt]
\centering
\topinset{(a)}{
\includegraphics[width=6.5cm]{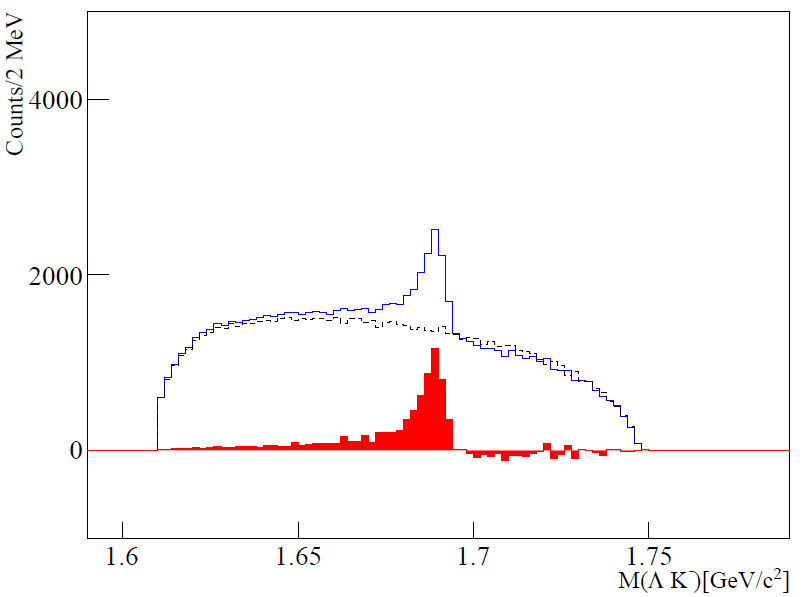}
}{0.5cm}{1.8cm}\hspace{-0.4cm}
\topinset{(b)}{ 
\includegraphics[width=6.5cm]{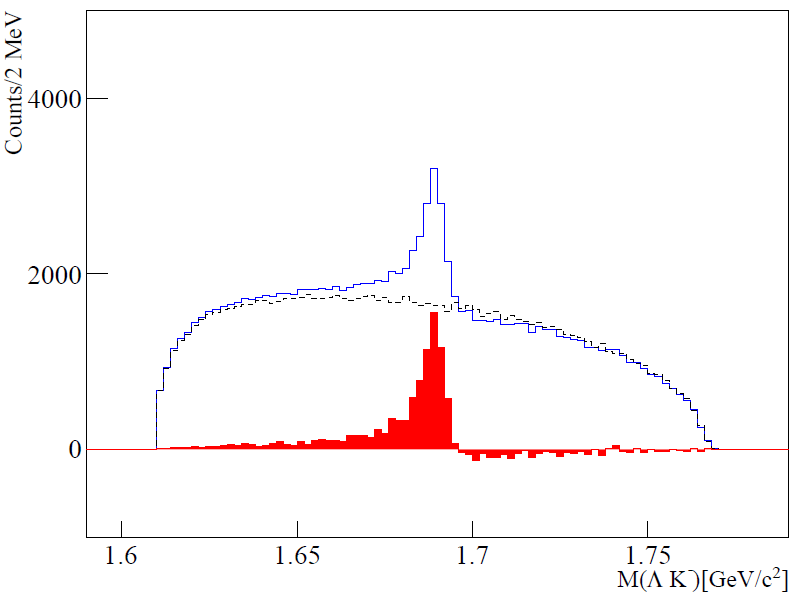}
}{0.5cm}{1.8cm} 
\caption{(Color online) Projected $\Lambda K^-$ mass distributions
for (a) $p^{\rm lab}_{K^-}=2.015$ GeV$/c$ and (b) $2.065$ GeV, respectively.}       
\label{fig:xi}
\end{figure}
%============================================================================

The calculated Dalitz plots show that we can neglect the interference effect between the $\Xi(1690)^-$ 
and the $\phi(1020)$ production channels in the $K^-p\to K^+\Xi(1690)^-$ reaction. However, it is
interesting to see that our theoretical model calculation predicts
possible interference between
the $\Xi(1690)$ and tree-level Born-term amplitudes. Excluding the $\phi(1020)$ band, the projected
$\Lambda K^-$ mass distributions for the beam momenta $p_{K^-}=2.015$ and 2.065 GeV$/c$ are displayed in
Fig. \ref{fig:xi}(a) and (b), respectively. 
The lineshape of the $\Xi(1690)^-$ is clearly observed in the $\Lambda K^-$ mass
spectrum. The tree-level Born-term contribution is subtracted 
from the projected $\Lambda K^-$ mass distribution, as shown with overlaid red distributions. The
subtracted distributions are made of the $\Xi(1690)^-$ and the interference effect.  

%=======================================================
%Ver.2.1 
%=======================================================
Finally, we want to discuss the spin and parity of $\Xi(1690)^-$, 
which has not yet been fully determined experimentally, 
although the BaBar Collaboration reported that $J^P=1/2^-$ assignment was favored ~\cite{babar}. 
Note that the theoretical predictions also support $J^P=1/2^-$~\cite{xiao,khem}, which we have employed 
for the numerical results shown above. For other possible spin-parity states, we use the following branching ratios
suggested by the ChUM calculations~\cite{khem}, which are qualitatively consistent with experimental results~
\cite{babar}:
%EQUATION>>>
\begin{equation}
\label{eq:DECAY}
\frac{\Gamma_{\Xi(1690)\to\overline{K}\Lambda}}{\Gamma_{\Xi(1690)}}=0.271,\,\,\,\,
\frac{\Gamma_{\Xi(1690)\to\overline{K}\Sigma}}{\Gamma_{\Xi(1690)}}=0.533.
\end{equation}
%EQUAITON>>>
These values provide the relevant coupling constants as follows:
%EQUATION>>>
\begin{eqnarray}
\label{eq:SCCC}
g_{\bar{K}\Lambda\Xi(1690)}&=&-(2.38,\,\,0.90,\,\,8.50),
\cr
g_{\bar{K}\Sigma\Xi(1690)}&=&(20.5,\,\,7.33,\,\,344.8)
\end{eqnarray}
%EQUAITON>>>
for $J^P=(1/2^+, 3/2^+, 3/2^-)$, respectively. Here, we have chosen the phase factor $-1$ between the two couplings 
for brevity \cite{khem}. 
The cutoff masses for the form factors are taken as $\Lambda_{\Xi(1690)}=(440,~2400,~650)$ MeV, which  
fairly reproduce the $K^-p\to K^+K^-\Lambda$ data ~\cite{Flaminio:1983fw}.

%FIGURE-===========================================================================
 
\begin{figure}[!hpbt]
\centering
\topinset{(a)}{
\includegraphics[width=6.5cm]{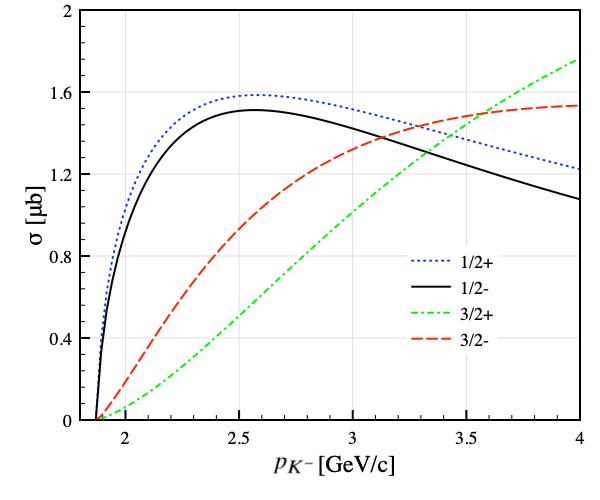}
}{0.5cm}{1.8cm}\hspace{-0.4cm}
\topinset{(b)}{ 
\includegraphics[width=6.5cm]{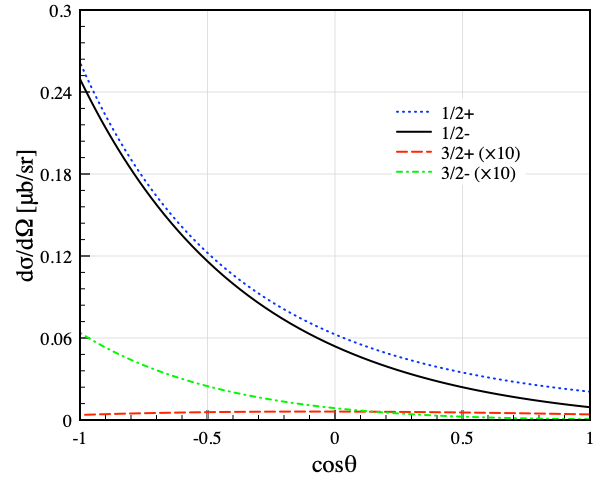}
}{0.5cm}{1.8cm} 
\caption{(Color online) Total cross sections for $K^-p\to K^+\Xi(1690)^-$ for different spin-parity states (a). Differential cross section for $p^{\rm lab}_{K^-}=2.015$ GeV$/c$ in the same manner (b).}       
\label{fig:COMP}
\end{figure}
%============================================================================

In Fig.~\ref{fig:COMP}(a), we plot the total cross sections for $K^-p\to K^+\Xi(1690)^-$ 
for different spin-parity states. The total cross sections for $J=1/2$ states increase rapidly near the threshold,
whereas those for $J=3/2$ states increase smoothly due to the $p$-wave nature near the threshold. 
The differential cross sections for $p^{\rm lab}_{K^-}=2.015$ GeV$/c$ in Fig.~\ref{fig:COMP}(b)
show a strong enhancement at backward $K^+$ c.m. angles because of the dominant $u$-channel contributions
for $J=1/2^+, 1/2^-$ and $3/2^-$ states. For $J=3/2^+$ state, 
it turns out that the $s$- and $u$-channel contributions compete strongly with each other. 

Taking into account the lack of experimental and theoretical information on those quantum numbers, 
it is crucial to investigate theoretically physical observables that do not depend much on theoretical uncertainties,
such as the form factors and coupling constants. One of the observables satisfying this criterion is 
\textit{double-polarization asymmetry} for the present case $K^-p\to K^+\Xi(1690)^-$, which reads:
%EQUATION>>>
\begin{equation}
\label{eq:PA}
\Sigma(s_R)=\frac{d\sigma_{\uparrow}/d\Omega-d\sigma_\downarrow/d\Omega}
{d\sigma_\uparrow/d\Omega+d\sigma_\downarrow/d\Omega}.
\end{equation}
%EQUAITON>>>
Here, the subscripts $(\uparrow,\downarrow)$ denote the proton-target polarizations along a quantization axis, 
when the $\Xi(1690)^-$ possesses a fixed spin state ($s_R$). 
As understood from Eq.~(\ref{eq:PA}), the phenomenological form factors and coupling constants, 
being multiplied to the invariant amplitudes, are canceled 
approximately between the numerator and denominator, minimizing those theoretical uncertainties. 

\begin{figure}[!hpbt]
\centering
\includegraphics[width=6.5cm]{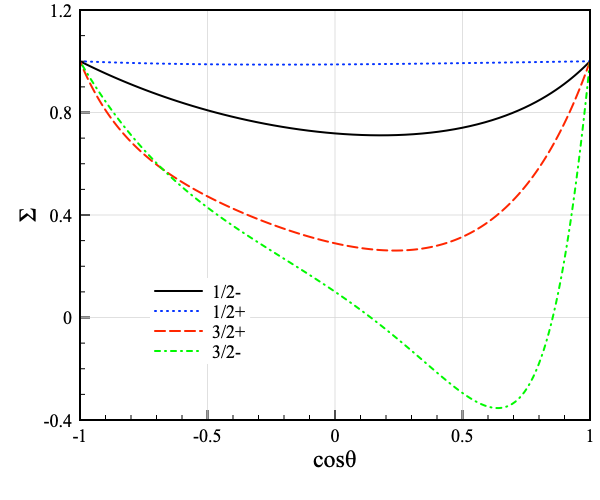}
\caption{(Color online) Double-polarization asymmetry $\Sigma(s_R)$ as a function of $\cos\theta$ for the various spin-parity states of $\Xi(1690)^-$ for $p^\mathrm{lab}_{K^-}=2.015$ GeV$/c$.}       
\label{fig:PA}
\end{figure}

In Fig.~\ref{fig:PA}, we plot $\Sigma(s_R)$ for different spin-parity states of $\Xi(1690)^-$ 
for $p^\mathrm{lab}_{K^-}=2.015$ GeV$/c$. We chose $J_z=+1/2$ for $J^P=1/2^\pm$ and summed over 
$J_z=+1/2$ and $J_z=+3/2$ contributions for $J^P=3/2^\pm$.
Because we do not have a meson exchange in the $t$ channel here, 
the spin-conserving process dominates for $J^P=1/2^\pm$, i.e., $\Sigma\sim 1$, as shown in Fig.~\ref{fig:PA}. 
Because the $J^P=1/2^-$ state provides sizable spin-nonconserving contributions, 
it differs slightly from the $J^P=1/2^+$ one. 
On the contrary, the $J^P=3/2^\pm$ states show both spin-nonconserving and spin-mixing contributions. 
Hence, from these observations, double-polarization asymmetry is a useful tool for determining 
the spin and parity quantum numbers of $\Xi(1690)^-$.
%=======================================================

\section{Summary}

In this study, we investigate the $\Xi(1690)^-$ production in the 
$K^-p\to K^+\Xi(1690)^-$ reaction within the effective Lagrangian approach. 
We consider the $s$- and $u$-channel $\Sigma/\Lambda$ ground states 
and resonances for the $\Xi$-pole contributions, 
in addition to the $s$-channel $\Lambda$, $u$-channel nucleon pole, 
and $t$-channel $K^-$-exchange for the $\phi$-pole contributions. 
The $\Xi$-pole includes $\Xi(1320)$, $\Xi(1535)$, $\Xi(1690)(J^p=1/2^-)$,
and $\Xi(1820)(J^p=3/2^-)$. We calculate the Dalitz plot density of 
$(d^2\sigma/dM_{K^+K^-}dM_{K^-\Lambda}$ at 4.2 GeV$/c$) and the total cross sections 
for the $K^-p\to K^+K^-\Lambda$ reaction near the threshold 
to determine the coupling constants and the form factors
for the two-body $K^-p\to K^+\Xi(1690)^-$ reaction.  
The calculated differential cross sections for the 
$K^-p\to K^+\Xi(1690)^-$ reaction near the threshold 
show a strong enhancement at backward $K^+$ angles, 
caused by the dominant $u$-channel contribution.
We also demonstrate that the Dalitz plot analysis for $p_{K^-}=1.915 - 2.065$ GeV$/c$ enables 
us to access direct information regarding the $\Xi(1690)^-$ production, 
which can be tested by future $K^-$ beam experiments. The double-polarization asymmetry 
turns out to be essential to determine the spin and parity quantum numbers of $\Xi(1690)^-$ via experiments.
%======================================================= 
%======================================================= 

\acknowledgments
S.i.N. is grateful to S.~H.~Kim (PKNU/CENuM) for fruitful discussions. 
%This research was also supported in part by 
%the Grant-in-Aid No. 14340075 for Scientific Research from the Ministry of Education, Culture, Sports, Science and 
%Technology (MEXT) Japan.
The works of J.K.A. and S.i.N. were partially supported by the National Research Foundation (NRF) of Korea 
(No. 2018R1A5A1025563, No. 2017R1A2B2011334).

%-------------------------------------------------


\begin{thebibliography}{99}
%-------------------------------------------------
\bibitem{pdg} 
  M.~Tanabashi {\it et al.}  (Particle Data Group),
  %``Review of Particle Physics,
  Phys.\ Rev.\ D {\bf 98}, 030001 (2018).
  %----------------------------------------------  
\bibitem{ramos} 
 A. Ramos, E. Oset and C. Bennhold,
  %Spin, Parity and Nature of the Xi(1620) Resonance
  Phys.\ Rev.\ Lett. {\bf 89}, 252001 (2002).
  %---------------------------------------------- 
  \bibitem{garcia} 
 C. Garcia-Recio, M. F. M. Lutz and J. Nieves,
  %
  Phys.\ Lett. \ B {\bf 582}, 49 (2004).
%----------------------------------------------  
\bibitem{oh}
Y. Oh,
Phys. Rev. D {\bf 75}, 074002 (2007).
%----------------------------------------------------
\bibitem{xiao} 
 L.~Y.~Xiao and X.~H.~Zhong,
  %``$\Xi$ baryon strong decays in a chiral quark model,
  Phys.\ Rev.\ D {\bf 87}, 094002 (2013).
%--------------------------------------------------
\bibitem{sekihara} 
  T.~Sekihara,
  %``$\Xi (1690)$ as a $\overline{K} \Sigma$ molecular state,
  PTEP {\bf 2015}, no. 9, 091D01 (2015).
%  doi:10.1093/ptep/ptv129
%  [arXiv:1505.02849 [hep-ph]].
%-------------------
\bibitem{miyahara} 
 K. Miyahara {\it et al.},
  % Theoretical study of the X(1620) and Xi(1690) resonances in X_c -> pi+MB decays
  Phys.\ Rev.\ C {\bf 95}, 035212 (2017).
  %-------------------
\bibitem{khem} 
 K. P. Khemchandani {\it et al.},
  %  
  Phys.\ Rev.\ D {\bf 97}, 034005 (2018).
%--------------------------------------------------
\bibitem{dionisi} 
  C.~Dionisi {\it et al.} (Amsterdam-CERN-Nijmegen-Oxford Collaboration),
  %``An Enhancement at the $\Sigma \overline{K}$ Threshold 1680-{MeV} Observed in $K^- p$ Reactions at 4.2-{GeV}/$c$,
  Phys.\ Lett.\ B {\bf 80}, 145 (1978).
%--------------------------------------------------
\bibitem{biagi}{S.F. Biagi {\it et al.}, {Z. Phys. C}, {\bf 9}, 305 (1981).}
\bibitem{biagi2}{S.F. Biagi {\it et al.}, {Z. Phys. C}, {\bf 34}, 15 (1987).}
\bibitem{belle}{K. Abe {\it et al.} (Belle Collaboration), Phys. Lett. B {\bf 524}, 33 (2002).} 
%--------------------------------------------------
\bibitem{babar} 
  B.~Aubert {\it et al.} (BaBar Collaboration),
  %``Measurement of the Spin of the Xi(1530) Resonance,''
  Phys.\ Rev.\ D {\bf 78}, 034008 (2008).
%  [arXiv:0803.1863 [hep-ex]].
%--------------------------------------------------
%  \bibitem{Barber:1975rg} 
%  P.~C.~Barber {\it et al.},
%  % T.~A.~Broome, B.~G.~Duff, F.~F.~Heymann, D.~C.~Imrie, G.~J.~Lush, E.~N.~Mgbenu and K.~M.~Potter {\it et al.},
%  %``K- p Elastic Scattering Between 1094-MeV/c and 1377-MeV/c,
%  Nucl.\ Phys.\ B {\bf 92}, 391 (1975).
%--------------------------------------------------
\bibitem{schlein} 
  P.~Schlein, W. E. Slater, L. T. Smith, D. H. Stork and H. K. Ticho,
  %Quantum numbers of a 1020-MeV KK resonance
  Phys.\ Rev.\ Lett. {\bf 10}, 368 (1963).
%--------------------------------------------------
\bibitem{badier}{J. Badier {\it et al.}, {Phys. Lett.}, {\bf 16}, 171 (1965).}
%--------------------------------------------------
\bibitem{bellefon} 
  A.~de Bellefon {\it et al.},
  %``Channel Cross-Sections of K- p Reactions from 1.934-GeV/c to 2.516-GeV/c,
  Nuovo Cim.\ A {\bf 41}, 451 (1977).
%-----------------------------------------
%--------------------------------------------------
\bibitem{gay} 
  J.~B.~Gay {\it et al.} (Amsterdam-CERN-Nijmegen-Oxford Collaboration),
  %``Production and Decay of xi* (1820) in K- p Reactions at 4.2-GeV/c,
  Phys.\ Lett.\ B {\bf 62}, 477 (1976).
%-------------------------------------------------
%--------------------------------------------------
\bibitem{Gasser:1984gg} 
  J.~Gasser and H.~Leutwyler,
  %``Chiral Perturbation Theory: Expansions in the Mass of the Strange Quark,
  Nucl.\ Phys.\ B {\bf 250}, 465 (1985).
%--------------------------------------------------
\bibitem{Lindsey:1966zz} 
  J.~S.~Lindsey and G.~A.~Smith,
  %``Production Properties and Decay Modes of the phi Meson,
  Phys.\ Rev.\  {\bf 147}, 913 (1966).
%--------------------------------------------------
\bibitem{Shyam:2011ys} 
  R.~Shyam, O.~Scholten and A.~W.~Thomas,
  %``Production of a cascade hyperon in the K^- - proton interaction,
  Phys.\ Rev.\ C {\bf 84}, 042201 (2011).
%  doi:10.1103/PhysRevC.84.042201
%  [arXiv:1108.2318 [hep-ph]].
%--------------------------------------------------
\bibitem{Nakayama:2006ty} 
  K.~Nakayama, Y.~Oh and H.~Haberzettl,
  %``Photoproduction of Xi off nucleons,
  Phys.\ Rev.\ C {\bf 74}, 035205 (2006).
%  doi:10.1103/PhysRevC.74.035205
%  [hep-ph/0605169].

%--------------------------------------------------
\bibitem{Rijken:1998yy} 
  T.~A.~Rijken, V.~G.~J.~Stoks and Y.~Yamamoto,
  %``Soft core hyperon - nucleon potentials,''
  Phys.\ Rev.\ C {\bf 59}, 21 (1999).
%  doi:10.1103/PhysRevC.59.21 [nucl-th/9807082].
%--------------------------------------------------
\bibitem{Nam:2004ub} 
  S.~i.~Nam, A.~Hosaka and H.~C.~Kim,
  %``Production of the pentaquark exotic baryon Xi(5) in anti-K N scattering: anti-K N ---> K Xi(5) and anti-K N ---> K* Xi(5),''
  J.\ Korean Phys.\ Soc.\  {\bf 52}, 561 (2008).
%  doi:10.3938/jkps.52.561  [hep-ph/0405227].
%--------------------------------------------------
\bibitem{Flaminio:1983fw} 
  V.~Flaminio, W.~G.~Moorhead, D.~R.~O.~Morrison and N.~Rivoire,
  %``Compilation Of Cross-sections. 2. K+ And K- Induced Reactions,''
  CERN-HERA-83-02.
%--------------------------------------------------
\bibitem{ahn}{J. K. Ahn, JPS Conf. Proc. {\bf 17}, 031104 (2017).}
\bibitem{ryu}{S. Y. Ryu, Letter of Intent for Study of Odd-Parity $\Sigma$ and $\Xi$ Resonances in $K^-p$ Reactions 
with the Hyperon Spectrometer at J-PARC (2015).}
\end{thebibliography}
\end{document}